\documentstyle[12pt,psfig]{article}
\def\reference{\parskip 0pt\par\noindent\hangindent 0.5 truecm}

\begin{document}
%
%
\title{Probing ISM Models with H$\alpha$ observations}
%


\author{ANDREA FERRARA $^{1}$
} 

\date{}
\maketitle

{\center
$^1$ Osservatorio Astrofisico di Arcetri, Largo. E. Fermi 5, Firenze, Italy,
50125 \\ ferrara@arcetri.astro.it \\ [3mm]
}

%
\begin{abstract}
I review the capabilities of H$\alpha$ observations to constrain some aspects
of the current models of the interstellar medium. In particular, it is shown
that turbulence is a necessary ingredient of any viable model, since most 
of the energy produced by supernova explosions and ionizing radiation is 
stored in kinetic form
in the ISM. Various forms of turbulent energy dissipation, including cloud
collisions, are analyzed. Two additional aspects, concerning the existence
of galactic fountains and their relation with High Velocity Clouds,
and the extended ionized layer of spiral galaxies are discussed; some crucial
experiments are suggested.  
\end{abstract}


\bigskip

%
%

\section{Relevance of H$\alpha$ observations for the ISM}

In spite of the impressive observational and theoretical advances
occured in the last decade, there are still a number of important
issues concerning the structure and evolution of the Galactic interstellar
medium (ISM) which require more study. Among these, three particularly
relevant issues are discussed below.
First, one would like to know to which extent the energy budget
of the ISM is regulated by turbulence. Energy in this form is certainly
supplied to the gas by the concurrent action of supernova explosions,
winds from massive stars and HII regions. However, it is a difficult problem
to ascertain the relative power of these sources and the total amount
of energy compared to the one stored in different forms (i.e. thermal,
radiative, magnetic and cosmic ray energy). Next, we understand that 
the disk of the Milky Way is not a close system but it is connected to
the Galactic halo by different forms of energy transfer involving mass
entrainment (particularly hot gas but also cold gas and dust) and photons 
leaking from the production regions. It it therefore important, both for 
energetic and evolutionary (chemical and dynamical)
considerations to assess the relevance of these disk/halo interactions,
which, in brief, are often referred to as the Galactic Fountain.
Finally, after more than 10 years from its firm discovery, the mystery
of the existence of a vertically extended, ionized gas distribution, 
the so-called "Reynolds layer'' has not yet been dispelled.
The aim of this work is to show how $H\alpha$ observations can help us 
in making progresses in each of these areas. 
  

\section{Turbulent Models of the ISM}

Massive stars are probably the most important energy sources for the ISM. They
inject power both in radiative (with ionizing photons creating HII regions) and   
mechanical (supernova explosions) form. The rate of {\it kinetic} energy density 
deposited via photoionization is 
\begin{equation}
\label{hii}
W_u^{(k)}=\alpha^{(2)}\langle n_e^2\rangle\bar E_2\sim 1.9\times 10^{-25}
\left({\langle n_e^2\rangle\over 0.1 {\rm ~cm}^{-6}}\right){\rm~ ergs~
cm}^{-3}{\rm s}^{-1};
\end{equation}
the analogous quantity for a supernova explosion is
\begin{equation}
\label{sne}
W_s^{(k)} \sim 2.2\times 10^{-25} \left({\gamma\over 0.04{\rm ~yr}^{-1}}\right)
\left({V_G\over 78  {\rm ~kpc}^{3}}\right)^{-1}{\rm~ ergs~
cm}^{-3}{\rm s}^{-1},
\end{equation}
where $\gamma$ is the total (both Type I and II) supernova rate and $V_G$ is the
volume of the Galaxy. Assuming an efficiency of kinetic energy conversion into the
ISM  $\eta_u \sim 1\%$ and $\eta_s \sim 3\%$ for radiative and mechanical input,
respectively, we find the total rate per unit volume at which energy 
available for motions in the ISM is produced:              
$W_i^{(k)} = \eta_u W_u^{(k)} + \eta_s W_s^{(k)} \sim 8\times 10^{-27}$~ergs~
cm$^{-3}$~s$^{-1}$. Kinetic energy is mostly dissipated by cloud collisions
at a rate
\begin{equation}
\label{ccc}
W_c^{(k)} \sim 2.2\times 10^{-27} n \left({v_r\over 14 {\rm ~km~s}^{-1}}\right)
\left({t_i\over 1.1\times 10^7 {\rm ~yr}}\right)^{-1}{\rm~ ergs~
cm}^{-3}{\rm s}^{-1}\sim 0.25 W_i^{(k)}, 
\end{equation}
where $n, v_r$ and $t_i$ are the typical cloud density, relative velocity and the mean time
interval between collisions, respectively. Thus, there is clearly enough energy 
production to support the observed motions in the ISM. 

\begin{figure}
 \begin{center}
 \psfig{file=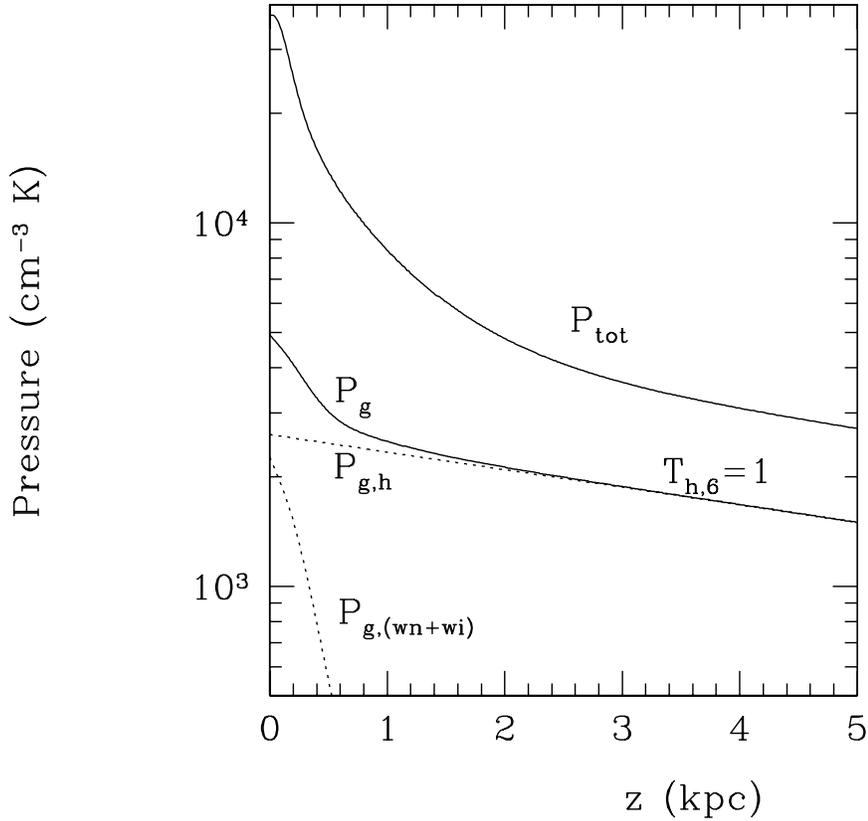,height=16cm}
 \caption{Thermal pressure contributions $P_{g,i}$ to the total pressure
 $P_{tot}$ (determined from hydrostatic equilibrium) from the known gaseous 
 phases of the ISM, where $i=wn,wi,h$ indicate warm neutral, warm ionized, 
 hot components, respectively. The difference $P_{tot}-P_g$ shows the importance
 of non-thermal (turbulent, magnetic, cosmic ray) forms of energy density.}
 \label{pressure}            
 \end{center}
 \end{figure}

These bulk motions have a strong impact on the structure of the Galactic
ISM. For example, they are shown to largely regulate and reproduce 
the vertical distribution 
of the HI in the gravitational potential of the Galaxy (Lockman \& Gehman 1991), 
once the effect of radiation pressure on dust grains embedded in clouds (the
so-called ``photolevitation'', Ferrara 1993) is properly taken into account.
In addition, turbulence may be the most important form of energy storage in
the ISM, as can be appreciated from Fig. \ref{pressure}: the thermal pressure 
contributed by all known gaseous ISM phases, $P_g$, appears to be at most
13\% of the total pressure, $P_{tot}$, as derived by imposing gas 
hydrostatic equilibrium
in the galactic gravitational field. Such large ratio of the turbulent to
thermal pressure, implies a large porosity factor, $Q$, of the hot gas. 
McKee (1990) estimates that $Q \sim 1.1 (P_{turb} / P_{tot})^{4/3}$, implying
$Q=0.92$ from the above estimates if the non-thermal energy is predominantly in
turbulent form; this corresponds to a hot gas filling factor  $f \sim 0.6$.

\begin{figure}
 \begin{center}
 \psfig{file=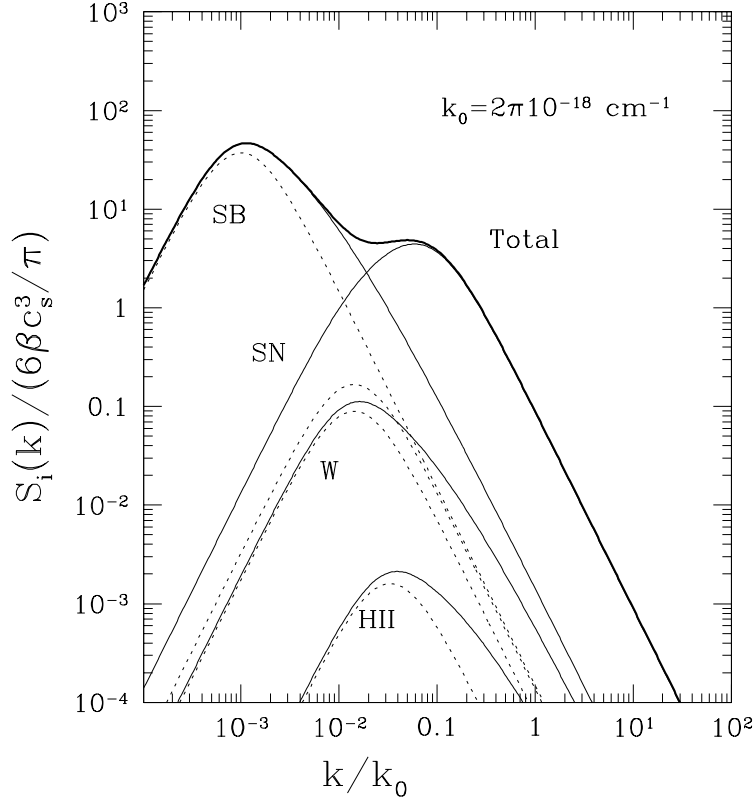,height=16cm}
 \caption{Normalized turbulent source functions $S_i(k)$ for supernovae,
 superbubbles, winds and HII regions as a function of normalized wavenumber. 
 {\it Solid lines} show the sum of primary
 and secondary shock contributions for each source; {\it dashed lines} show
 secondary shocks only (see Norman \& Ferrara 1996). The thick line is the 
 total grand source function.}
 \label{source}              
 \end{center}
 \end{figure}

Given these arguments it seems necessary to revise the current ISM models to
include turbulence. A first attempt in this direction has been carried out by
Norman \& Ferrara (1996). The authors calculate the detailed grand source
function (shown in Fig. \ref{source}) for the conventional sources of turbulence 
from supernovae, superbubbles, stellar winds and HII regions. As seen from Fig.
\ref{source}, superbubbles are the main contributors to interstellar turbulence.
In addition, from the study of the general properties of the
turbulent spectrum using an approach based on a spectral
transfer equation derived from the hydrodynamic Kovasznay
approximation, they conclude that the turbulent pressure calculated 
from the grand source function is $P_{turb} \sim 10-100~P_{g}$. Also,
given the scale dependent
energy dissipation from a turbulent cascade, the multi-phase medium
concept has to be generalized to a more natural continuum description
where density and temperature are functions of scale. 

\begin{figure}
 \begin{center}
 \psfig{file=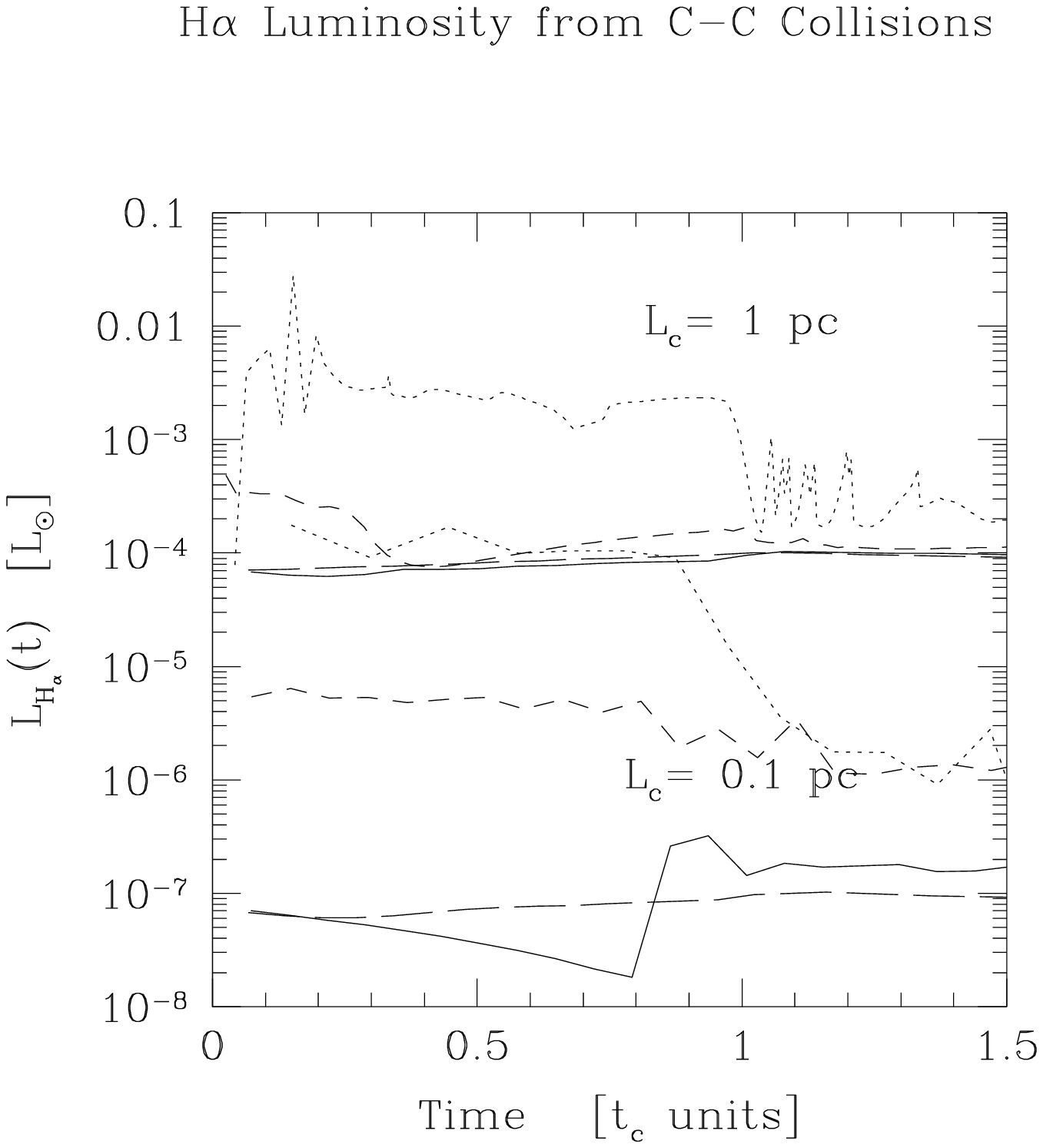,height=16cm}
 \caption{H$\alpha$ luminosity evolution from cloud collisions at a distance of 1 kpc 
 as a function of time in units of collision time ($t_c=L_c/v_r$). 
 The upper (lower) group of lines refer to a cloud size $L_c=1$~pc ($L_c=0.1$~pc) 
 for different values of the relative velocity of the collision:
 $v_r=6$~km~$s^{-1}$ ({\it solid}), $v_r=16$~km~$s^{-1}$ ({\it long-dashed}),
 $v_r=31$~km~$s^{-1}$ ({\it short-dashed }), $v_r=63$~km~$s^{-1}$ ({\it dotted})}
 \label{halfa}               
 \end{center}
 \end{figure}

As recalled above, cloud collisions represent the most efficient dissipation 
mechanism of large scale turbulence. The simple estimate for $W_c^{(k)}$ given
above, assumes that {\it all} the kinetic energy of the clouds is radiated away 
by the post-shock gas, i.e. an inelastic collision. This hypothesis is correct
only in a restricted region of the collision parameters (velocity and mass
ratio of the colliding clouds, magnetic field strength, gas metallicity).    
Ricotti etal. 1997 have studied the dependence of the elasticity (defined as the
ratio of the final to the initial kinetic energy of the clouds) on such
parameters (recently extended to include pre-interaction with the intercloud
medium by Miniati etal. 1997). They find that $(i)$  the collision elasticity is maximum
for a cloud relative velocity $v_r \simeq 30$ km s$^{-1}$;
$(ii)$ the elasticity is  $\propto Z L_c^2$, where $Z$ is the
metallicity and $L_c$ is the cloud size: the larger is $ZL_c^2$, the
more dissipative (inelastic) the collision will be. 
During the collision the warm post-shock gas  will radiate a substantial
fraction of its internal energy in the H$\alpha$ line depending on 
$v_r$ and $L_c$. Fig. \ref{halfa} shows the H$\alpha$ 
luminosity for a collision occurring $\sim 1$~kpc away from us for different
values of $v_r$ and $L_c$. Thus, H$\alpha$ observations can be used in principle
as a powerful indicator of large scale turbulent motion dissipation. 
  

\section{Does a Galactic Fountain exist ?}

\begin{figure}
 \begin{center}
 \psfig{file=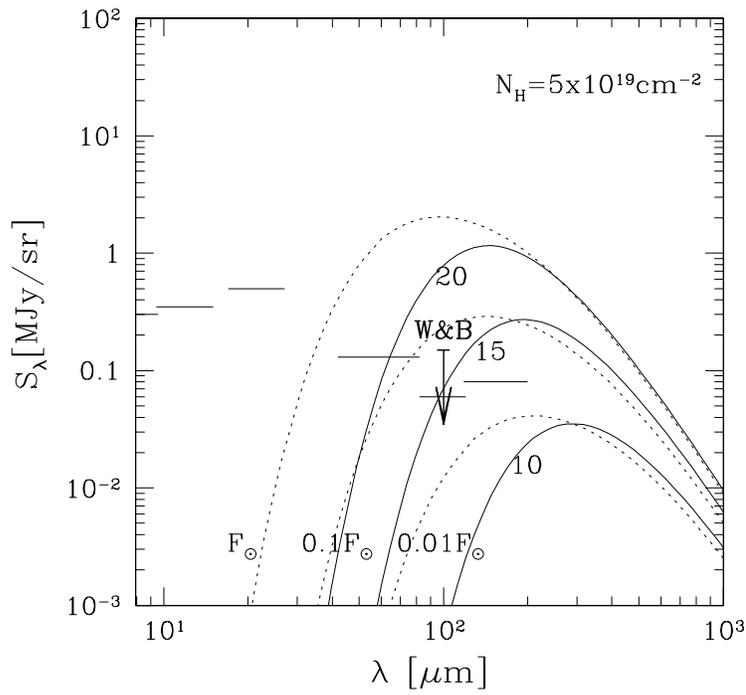,height=16cm}
 \caption{IR surface brightness from dust in HVCs with HI column density
 $N_{HI}=5\times 10^{19}$~cm$^{-2}$. Numbers refer to fixed grain temperatures
 ({\it solid})) or radiation flux intensity ({\it dotted}),
 where $F_\odot$ is the Galactic ISRF. Also shown are the
 Wakker \& Boulanger limits and ISOPHOT sensitivity (see text) }
 \label{dust}                
 \end{center}
 \end{figure}

One of the major predictions and elements in favor of the Galactic Fountain (GF)
model (Shapiro \& Field 1976, subsequently detailed by Bregman 1980) is the
existence of the High Velocity Clouds (HVCs, recently reviewed by Wolfire etal. 1995) 
in the halo, as a by-product of the cooling of the hot fountain gas.
Any constraint on the origin of HVCs would be highly valuable in terms
of understanding the global disk/halo circulation. Ferrara \& Field (1994)
have investigated the ionization and thermal structure of  HVCs   
due to the extragalactic background radiation field and calculated the 
H$\alpha$ emission from the partially ionized edge of a given cloud.
Comparing the model results to the available H$\alpha$ observations,  
the authors found that the observed H$\alpha$ intensity is larger
than the predicted one for all the different cases considered. One  
way to reconcile this dicotomy is that some of the observed
emission is H$\alpha$ light coming from the disk of the Galaxy and 
subsequently back-scattered by dust in the clouds. Assessing the presence of dust
in HVCs is a young research field and very little is known about this issue.  
However, there are theoretical bases to expect dust to be present. 
The efficiency of dust destruction in a shock (presumably the heating source
for the fountain gas) is likely to be less than 10\% (McKee 1989); the 
subsequent thermal sputtering in the hot gas might destroy some of the smallest
grains: in a $10^6$~K gas, the sputtering time is shorter than the gas 
cooling time only for grains sizes smaller than $0.03 \mu$m. However, previous
low sensitivity searches using IRAS data have produced only upper limits (Wakker
\& Boulanger 1986) on HVCs dust content. 
In addition to the unlikely complete dust depletion, there
are at least two other possibilities which might explain the non-detection: 
(i) HVCs are far above the galactic plane ($\ge 10$~kpc); (ii) the
dust is too cold to emit substantially in the IRAS bands. Since, at least for 
some of the clouds, the distance is bracketed around  a much lower value for
the distance, (i) is an unlikely explanation. It is quite possible, instead,
that the dust is cold, due to the very diluted halo radiation field. The
expected IR emission from a $N_{HI}=5\times 10^{19}$~cm$^{-2}$ HVC,  
assuming either a given temperature 
of the grains or a galactic ISRF diluted by a given factor (1-100 times) 
is shown in Fig. \ref{dust} together with ISOPHOT surface brightness limits
(128 s of integration time, S/N=10) in the various bands. Clearly, dust hotter
than $\sim 10$~K and/or heated by a ISRF diluted by at most 10 times, could be
detectable. Such a detection would bring strong support to the idea that HVCs 
are not of intergalactic origin and they are very likely part of a global
disk/halo circulation.
  

\section{H$\alpha$ from the galactic DIG}

The thick ($z\sim 1 $~kpc) {\it Reynolds layer} of diffuse ionized gas (DIG)
discovered in the Galaxy and in external ones poses some of the most challenging
problems for our understanding of the large scale structure of the Galactic ISM
(Reynolds 1995). One of the best ways to study this component is represented by
spectroscopic observations of emission lines like  $H\alpha$, $[NII](\lambda
6583$\AA)
and $[SII](\lambda 6716$\AA).  The different excitation conditions found
in a given galaxy as a function of height above the plane pose a relevant
question concerning the amount of light originating in the disk (where the most
obvious
ionization sources are located) and light scattered back into the line of sight
by dust. This aspect
has been investigated recently by Ferrara etal. (1996),  using Monte Carlo
simulations to calculate the  radiation
transfer of $H\alpha$ line emission, produced both by HII regions in the
disk  and in the diffuse ionized gas (DIG), through the dust layer of the
galaxy NGC891.
  
\begin{figure}
 \begin{center}
 \psfig{file=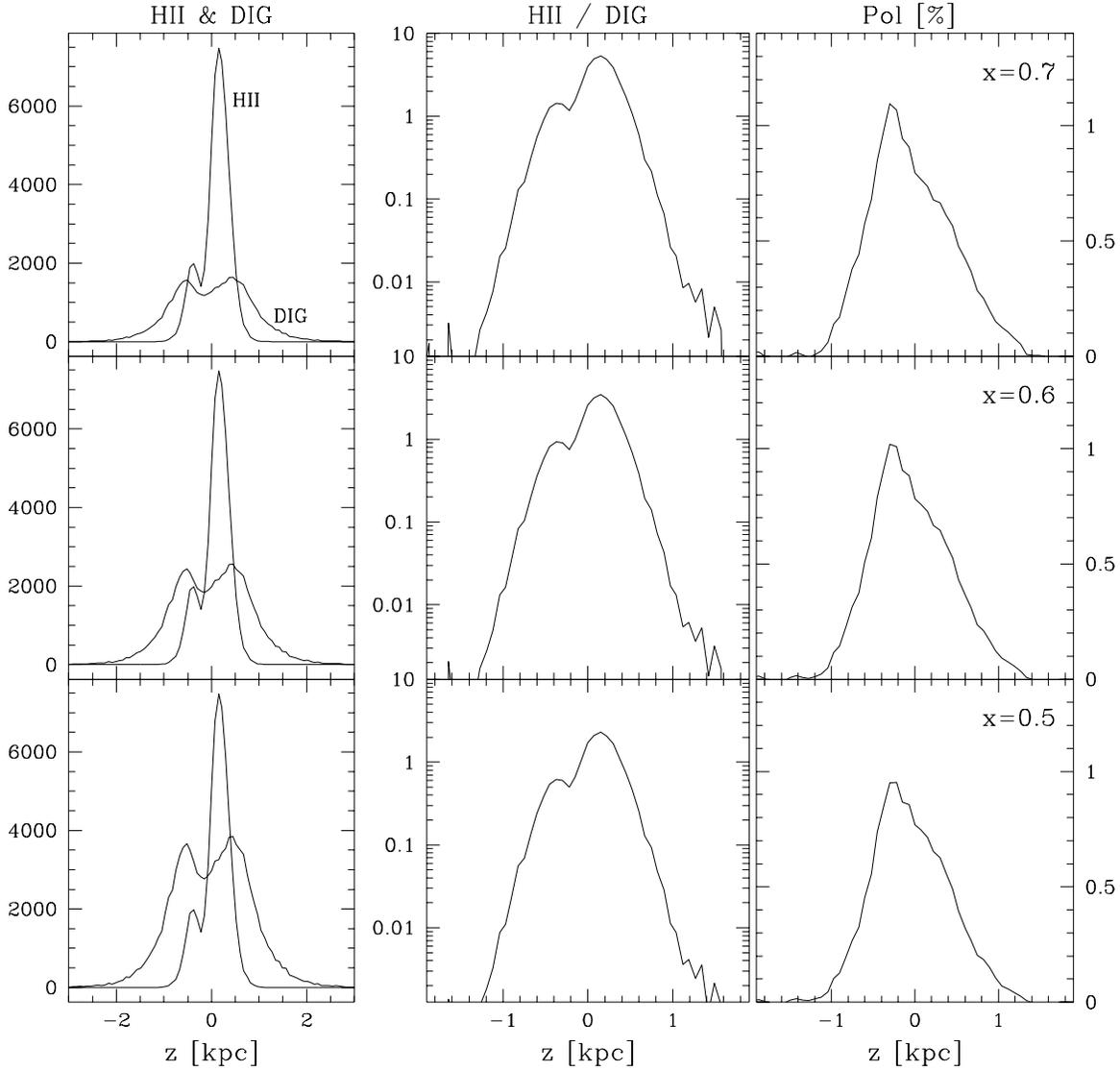,height=16cm}
 \caption{$z$-axis cuts through the center of NCG891
 for (left column) $H\alpha$ luminosity profiles corresponding to contributions
 of HII regions and DIG component, (middle) ratio of the two components, and
(right) profiles of the linear polarization degree.  Plots refer to the cases of
 $x$=0.7, 0.6, 0.5 (from top to bottom) where $x$ is the ratio between $H\alpha$
 luminosity from HII regions to the total (DIG + HII regions) one.}
 \label{scatter}
 \end{center}
 \end{figure}

The amount of light originating in the HII
regions of the disk and scattered by EGD can be then compared
with the emission produced by recombinations in the DIG.
The cuts of photometric and polarimetric maps along the $z$-axis
show that scattered light from HII regions is still 10\% of
that of the DIG at $z\sim 600$~pc (Fig. \ref{scatter}), whereas the
degree of  linear polarization is small ($<1$\%).
This could explain the observed behavior of emission line ratios as a function
of height (Ferrara etal. 1996).

%
%





\section*{Acknowledgements}

I deeply thank all my collaborators that have been part of this project:
S. Bianchi, R. Dettmar, G. Field, C. Norman and M. Ricotti

\section*{References}






\reference  Bregman, J. N. 1980, ApJ, 236, 577 

\reference  Ferrara, A. 1993, ApJ, 407, 157

\reference  Ferrara, A. \& Field, G. B. 1994,  ApJ, 423, 665

\reference  Ferrara, A., Bianchi, S.,  Dettmar, R.-J., \& Giovanardi, C. 1996, ApJL, 
467, 69

\reference  Lockman, F.J., \&  Gehman, C.S. 1991,  ApJ, 382, 182

\reference McKee, C. F. 1989, in Interstellar Dust, IAU Symp. 135, eds. 
Allamandola L. J. \& Tielens A. G. G. M.,  Kluwer, 431 

\reference  McKee, C.F. 1990, in The Evolution of the Interstellar Medium,
	 ed. Blitz L., PASP, 3

\reference  Miniati, F., Jones, T. W., Ferrara, A. \& Ryu, D. 1997, ApJ,
submitted (astro-ph/9706208)

\reference Norman, C. A., \& Ferrara, A. 1996, ApJ, 467, 280

\reference Reynolds, R. J. 1995, in The Physics of the Interstellar Medium and
Intergalactic Medium, eds. A. Ferrara etal., ASP Conf. Series, 80, 388

\reference Ricotti, M., Ferrara, A. \& Miniati, F. 1997, ApJ, in press 
(astro-ph/9702143)

\reference Shapiro, P. R. \& Field, G. B. 1976, ApJ, 205, 762

\reference Wakker, B. P. \& Boulanger, F 1986, A\&A, 170, 84

\reference Wolfire, M.G., Hollenbach, D., McKee, C.F., Tielens, A.G.G.M., \&
Bakes, E.L.O. 1995, ApJ, 443, 152
  
\end{document}